\documentclass[aps,twocolumn,pra,floatfix]{revtex4}
\usepackage{epsfig}
\usepackage{graphicx}
\usepackage{dcolumn}
\usepackage[colorlinks,allcolors=blue]{hyperref}

\begin{document}
	
\title{Effects of the long-range neutrino-mediated force in atomic phenomena}
\author{ P. Munro-Laylim, 
    V. A. Dzuba, 
and V. V. Flambaum 
}
\affiliation{School of Physics, University of New South Wales, Sydney 2052, Australia}

\begin{abstract}
As known, electron vacuum polarization by nuclear Coulomb field produces Uehling potential with the range $\hbar/2m_e c$. Similarly, neutrino vacuum polarization  by $Z$ boson field produces long range potential $\sim G_F^2/r^5$  with a very large range $\hbar/2m_{\nu}c$. Measurements of  macroscopic effects produced by  potential  $G_{\rm{eff}}^2/r^5$ give limits on the effective interaction constant  $G_{\rm{eff}}$ which exceed Fermi constant $G_F$ by many orders of magnitude, while limits from spectroscopy of simple atomic systems are  approaching the Standard Model predictions. In the present paper we consider limits on $G_{\rm{eff}}$ from hydrogen, muonium, positronium, deuteron, and molecular hydrogen. Constraints are also obtained on fifth force parameterised by Yukawa-type potential mediated by a scalar particle.
\end{abstract}
	
\maketitle

\section{Introduction}\label{SectionIntroduction}
\begin{figure}[htb]
    \includegraphics[width=0.25\textwidth]{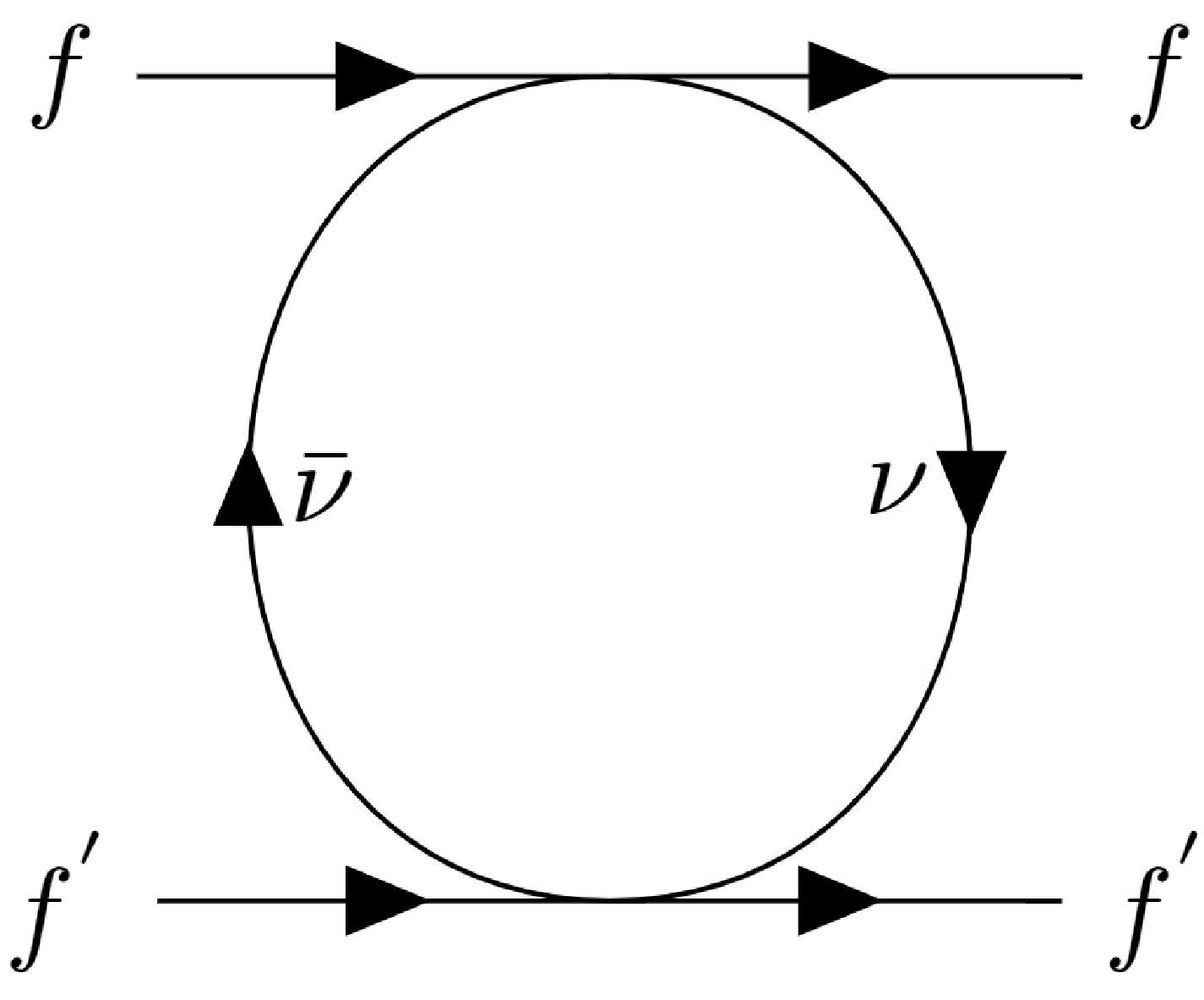}
    \caption{\label{fig:neutrino_exchange_feynmann} Diagram describing neutrino-exchange potential $\sim G_F^2/r^5$ based on Fermi-type  four-fermion  interactions.}
\end{figure}

It has long been known that the exchange of a pair of (nearly) massless  neutrinos between particles (see  diagram on Fig. \ref{fig:neutrino_exchange_feynmann}) produces a long-range force \cite{GamowPR1937,Feynman}, with the resultant potential $\sim G_F^2/r^5$, where $G_F$ is Fermi constant ~\cite{FeinbergPR1968,FeinbergPR1989,HsuPRD1994}. However, due to a rapid decay with the distance $r$, the  effects of this potential are about 20 orders of magnitude smaller than the  sensitivity of the  macroscopic experiments Refs.~\cite{Kapner2007,Adelberger2007,Chen2016,Vasilakis2009,Terrano2015,StadnikPRL2018}.

A recent paper by Stadnik \cite{StadnikPRL2018} introduced a new approach to obtaining constraints on this potential  by considering spectra of atomic systems.  In the Standard Model formulas  for energy shifts produced by potential $G_F^2/r^5$, the Fermi constant $G_F$ has been replaced by an effective interaction constant $G_{\rm{eff}}$. The $G_{\rm{eff}}^2/r^5$ potential produces a small
energy shift to atomic energy levels, 
and therefore it is possible to obtain constraints on   $G_{\rm{eff}}^2$ from differences between highly accurate QED calculations of energy levels and experimental results \cite{MuoniumMeyer2000,PositroniumKarshenboim2000}. The Stadnik  paper has lead to a breakthrough in sensitivity, constraints on the interaction constant $G_{\rm{eff}}^2$  have been improved by 18 orders of magnitude in comparison with constraints from the macroscopic  experiments Refs.~\cite{Kapner2007,Adelberger2007,Chen2016,Vasilakis2009,Terrano2015,StadnikPRL2018}.

However, the highly singular  potential $G_{\rm{eff}}^2/r^5$ leads to divergent integrals in the matrix elements as $r$ approaches zero. This  demonstrates the requirement of the correct extension of the potential for $r\to 0$. Ref. \cite{StadnikPRL2018} used the  Compton wavelength of the $Z$ boson as the cut-off radius, $r_c = \hbar/M_Z c$, for positronium and muonium, and the nuclear radius $R$
for atoms with finite nuclei. As we will show below, this oversimplified  treatment in  Ref. \cite{StadnikPRL2018} leads  to limits which were 
overestimated by a factor of 6 in non-hadronic atoms and underestimated by 4-5
orders of magnitude in the case of 
deuteron binding energy. The aim of the present paper is to provide more accurate estimates and also consider results of the measurements which have not been included in Ref. \cite{StadnikPRL2018}.  To avoid misunderstanding, we should note that present paper is not  aimed to calculate all electroweak  corrections to energy levels. This should be done by a different method. 

We also consider fifth forces from beyond the Standard Model that are parameterised by a Yukawa-type potential. This fifth force would require the existence of a new scalar particle to mediate the interaction, thus constraints on the coupling strength of the interaction can be found for various scalar particle masses. Limits were previously obtained using precision hydrogen $1s-2s$ spectroscopy in Ref. \cite{Yukawa_H}, however we improve upon them using more recent data and include additional hydrogen-like systems.

\section{The Long-Range Neutrino-Mediated Potential}\label{SectionPotential}

The potential of the long-range neutrino-mediated force between two particles, presented in Ref. \cite{StadnikPRL2018}, is
\begin{eqnarray}
    V_\nu(r) =&&\ \frac{G_F^2}{4\pi^3 r^5} \bigg( a_1 a_2 - \frac{2}{3} b_1 b_2 \mathbf{\sigma_1 \cdot \sigma_2}\nonumber\\
    &&\ -\frac{5}{6} b_1 b_2 [\mathbf{\sigma_1 \cdot \sigma_2} - 3 (\mathbf{\sigma_1 \cdot \hat{r}}) (\mathbf{\sigma_2 \cdot \hat{r}})] \bigg),
    \label{e:neutrino_ex_potential}
\end{eqnarray}
where 
$\sigma_1$ and $\sigma_2$ are the Pauli spin matrix vectors of the two particles, and $a_i$ and $b_i$ represent the species-dependent parameters defined below. It is worth noting that the last term of Eq.~(\ref{e:neutrino_ex_potential}) is zero for $s$-orbitals which strongly  dominate in the  shifts of atomic energy levels.

\begin{figure}[htb]
    \includegraphics[width=0.25\textwidth]{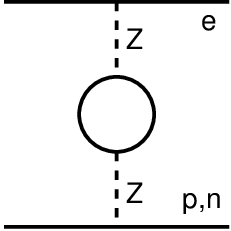}
    \caption{\label{fig:feynman-z-exchange} Vacuum polarization by the nuclear $Z$ boson field with a light fermion loop producing potential  with the range $\hbar/(2m c)$.  }
\end{figure}

A potential $\sim 1/r^5$ gives divergent integrals ($\int_{r_c} d^3r/r^5 \approx 1/2 r_c^2$) in the matrix elements for $s$-wave.  Using the nuclear radius $R$ as a cut-off, $r_c=R$,  would give incorrect results.  
A more accurate approach requires first to build effective potential for electron-quark interaction and then take into account nucleon distribution $\rho(r)$ inside the nucleus. To include small distances, we present this potential for the finite size $R$ of the nucleus and cut-off for large momenta (small distances $r$) produced by the $Z$ boson propagator ($1/(q^2 +M_Z^2$) instead of $1/M_Z^2$, see Fig. \ref{fig:feynman-z-exchange}). 
To start, we replace $1/r^5$ in the potential Eq.~(\ref{e:neutrino_ex_potential}) with
\begin{eqnarray}
    F(r) = \frac{8 m^4 c^4}{3 \hbar^4} \frac{I(r)}{r},
\label{e:F}\end{eqnarray}
where, for $z=M_Z/(2m)$,
\begin{eqnarray}
    I(r)= \int_1^{\infty} e^{-2x m c r/\hbar} \left(x^2- \frac{1}{4}\right) \frac{ \sqrt{x^2-1}  z^4 dx}{(x^2 +z^2)^2}.
\label{e:I}
\end{eqnarray}
Here $m$ is the mass of the fermion in the loop on  Fig. \ref{fig:feynman-z-exchange}. The function $F(r)\propto I(r)/r$ gives us dependence of interaction between electron and quark (or electron and other point-like fermion) on distance $r$ between them. For $\hbar/(M_Z c) \ll r \ll \hbar/(m c)$, we obtain $F(r)=1/r^5$. In this area there is no change for potential Eq.~(\ref{e:neutrino_ex_potential}). For large $r \gg  \hbar/(m c)$, we have $F(r) \propto \exp(-2 m c r/\hbar )/r^{5/2}$. At small distance $r \ll  r_c=\hbar/(M_Z c)$, function $F(r) \propto (\ln r)/r$ and has no divergency integrated with $d^3 r$. Note that behaviour of the neutrino-exchange potential at small distance has been investigated in Ref.~\cite{Xu2022}. However, they do not study this potential in the Standard Model, they considered a new scalar particle instead of $Z$ boson.

Convergence of the integral  in the matrix elements on the distance $r \sim r_c=\hbar/M_Z c$ indicates that this interaction in atoms may be treated as a  contact interaction (see Fig. \ref{fig:neutrino_exchange_feynmann}). We can replace $F(r)$ by its contact limit, $F(r) \to C\delta({\bf r})$ 
\begin{equation}\label{C}
C= \int_0^{\infty} F(r) d^3r=\frac{\pi}{3}\frac{M_Z^2 c^2}{\hbar^2}.
\end{equation}
where we assume $z=M_Z/(2m) \gg 1$. Note that if we would assume potential $1/r^5$ with the cut-off $r_c = \hbar/M_Z c$, the result would be 6 times bigger:
\begin{equation}\label{Cprime}
C'=\int_{r_c}^{\infty} \frac{1}{r^5} d^3r= 2 \pi \frac{M_Z^2 c^2}{\hbar^2}.
\end{equation}
Using Eq.~(\ref{C}), the potential in Eq.~(\ref{e:neutrino_ex_potential}) in the contact limit may be presented as, using natural units $\hbar=c=1$,
\begin{eqnarray}
    V^{C}_\nu(r) =&& \frac{G_F^2 M_Z^2 \delta(\mathbf{r})}{12\pi^2}\bigg( a_1 a_2 - \frac{2}{3} b_1 b_2 \mathbf{\sigma_1 \cdot \sigma_2}\nonumber\\
    &&\ -\frac{5}{6} b_1 b_2 [\mathbf{\sigma_1 \cdot \sigma_2} - 3 (\mathbf{\sigma_1 \cdot \hat{r}}) (\mathbf{\sigma_2 \cdot \hat{r}})] \bigg)\nonumber\\
    \equiv&& g \delta(\mathbf{r}).
    \label{e:contact_neutrino_ex_potential}
\end{eqnarray}

In Ref. \cite{Grifols}, the potential was obtained for a Majorana neutrino loop instead of a Dirac neutrino loop. Using these results, we conclude that the neutrino-exchange potential for Majorana neutrinos requires the adjustment to $I(r)$ as follows
\begin{eqnarray}
    I^{(M)}_2(r) = \int_1^{\infty} e^{-2x m c r/\hbar} \frac{ (x^2-1)^{3/2}  z^4 dx}{(x^2 +z^2)^2}.
    \label{e:IM}
\end{eqnarray}

This indicates that the nature of neutrinos may, in principle,  be detected from the difference in Dirac and Majorana potentials. At small distance, the Dirac neutrino and Majorana neutrino potentials are practically the same, the difference is proportional to $(m_\nu c r/\hbar)^2$ and is very small. In the contact interaction limit, the relative difference is $\sim (m_\nu/M_Z)^2$. However, the asymptotic expression at large distance changes: for Majorana neutrinos we have  $I^{(M)}_2(r)/r \propto \exp(-2 m c r/\hbar )/r^{7/2}$, whereas $I(r)/r \propto \exp(-2 m c r/\hbar )/r^{5/2}$ for Dirac neutrinos. Therefore, the ratio of Dirac potential to Majorana potential $ \sim m_{\nu} c r/\hbar$ \cite{Grifols}.  Thus,  the difference is negligible at small distances and only becomes significant at large distances $r\gtrsim \hbar / m_{\nu } c$. Unfortunately,  effects of the neutrino-exchange potential are many orders of magnitude smaller than sensitivity of current macroscopic experiments Refs.~\cite{Kapner2007,Adelberger2007,Chen2016,Vasilakis2009,Terrano2015}, motivating future experimental work.

At large distance a dominating contribution to the vacuum polarization by the $Z$ boson field is given by the lightest particles which are neutrinos. However, at distance $r$ all particles with the Compton wavelength  $\hbar/m c > r$ give a significant contribution.
Following Ref.   \cite{StadnikPRL2018} we present interaction constants for potentials (\ref{e:neutrino_ex_potential},\ref{e:contact_neutrino_ex_potential}) in the following form:  
\begin{eqnarray}
    a_1 a_2 &=& a_1^{(1)} a_2^{(1)} + (N_{\rm{eff}} - 1) a_1^{(2)} a_2^{(2)},\label{a}\\
    b_1 b_2 &=& b_1^{(1)} b_2^{(1)} + (N_{\rm{eff}} - 1) b_1^{(2)} b_2^{(2)},\label{b}
\end{eqnarray}
where $N_{\rm{eff}}$ is the effective number of particles (normalised to  one  neutrino contribution) mediating the interaction on Fig. \ref{fig:feynman-z-exchange}. Contribution, which is not proportional to $N_{\rm{eff}}$, appears due the diagrams with $W$ boson. For example, for interaction between  electron  and quark, such diagrams involve  electron neutrino - see Ref. \cite{HsuPRD1994}.

In atoms dominating contribution comes from the distance $r\sim \hbar/M_Z c$. Summation of the contributions from $\nu, \, e,\,\mu,\, \tau, \, u,\, d,\, s,\, c,\, b$ (all with mass $m \ll M_Z$) gives $N_{\rm{eff}} = 14.5$ \cite{StadnikPRL2018}. 
Consider an interaction between electron and nucleon with an exchange by electron neutrino, electron has values $a_e^{(1)} = 1/2 + 2\sin^2(\theta_W)$ and $b_e^{(1)} = 1/2$,  while nucleons have values $a_n^{(1)} = -1/2$, $a_p^{(1)} = 1/2 - 2\sin^2(\theta_W)$, $b_n^{(1)} = -g_A/2$, and $b_p^{(1)} = g_A/2$, where   $g_A \approx 1.27$. For the contributions from the other neutrino species, there is no $W$ boson contribution and  we have values for charged leptons $a_l^{(2)} = 2\sin^2(\theta_W) - 1/2$, $b_l^{(2)} = -1/2$, $a_N^{(2)} = a_N^{(1)}$, and $b_N^{(2)} = b_N^{(1)}$.   Value of the $\sin^2(\theta_W)=0.239$ for a small  momentum transfer \cite{SM}, where $\theta_W$ is the Weinberg angle.

\section{Energy Shift in Hydrogen-Like Systems}\label{SectionSimpleSystems}
Simple two-body systems provide the most accurate values of the difference between experimental result  and result of QED calculation of the transition energies. Following Ref. \cite{StadnikPRL2018}, we use these differences to obtain limits on the effective interaction constant $G_{\rm{eff}}$. We consider hydrogen, muonium and positronium spectra and deuteron binding energy. A summary of our  calculations is presented in Table~\ref{tab:SimpleSystemResults}.

\subsection{Hydrogen Spectroscopy}
For a simple hydrogen-like system, the expectation value of a contact potential $g\delta(\mathbf{r})$ is
\begin{equation}\label{e:gen_expval}
    \langle\psi|g\delta(\mathbf{r})|\psi\rangle = \frac{g Z^3}{n^3 \pi \tilde{a}_B^3},
\end{equation}
where $Z$ is the atomic charge, $n$ is the principal quantum number, and $\tilde{a}_B$ is the reduced Bohr radius. Therefore, we calculate the energy shift for $n^3 S_1$ states in hydrogen using Eq.~(\ref{e:contact_neutrino_ex_potential}) and (\ref{e:gen_expval}), 
\begin{equation}
    \delta E_{n^3 S_1} = -\frac{G_F^2 M_Z^2 Z^3}{12\pi^3 n^3 \tilde{a}_B^3} \bigg(a_e a_p - \frac{2}{3} b_e b_p\bigg).
\end{equation}

The energy shift for hydrogen $1s-2s$ ($Z=1$ and $\tilde{a}_B = a_B$) evaluates to
\begin{equation}
    \delta E = 3.60\times10^{-16}\ \text{eV}.
\end{equation}
From Ref. \cite{H1s2s}, the maximal energy difference is $E_{\rm{exp}} - E_{\rm{thr}} = 2.2\times10^{-11}$ eV. For the neutrino-mediated potential to account for this maximal difference, we replace $G_F^2$ with the effective interaction constant $G_{\rm{eff}}^2$ to obtain constraints
\begin{equation}
    \delta E \left(\frac{G_{\rm{eff}}^2}{G_F^2}\right) \leq E_{\rm{exp}} - E_{\rm{thr}},
\end{equation}
\begin{equation}
   G_{\rm{eff}}^2 \leq 6.1\times10^4\ G_F^2.
\end{equation}

Similarly for hydrogen $1s-3s$, we find the energy shift to be
\begin{equation}
    \delta E = 3.97\times10^{-16}\ \text{eV}.
\end{equation}
Using the maximal energy difference is $E_{\rm{exp}} - E_{\rm{thr}} = 2.2\times10^{-11}$ eV from Ref. \cite{H1s3s}, we then get the constraint
\begin{equation}
   G_{\rm{eff}}^2 \leq 5.5\times10^4\ G_F^2.
\end{equation}

Both of these constraints are strong and were not included in Ref. \cite{StadnikPRL2018}.

\subsection{Muonium and Positronium 1\textit{s}-2\textit{s}}
We calculate the energy shift for $n^3 S_1$ states for muonium and positronium using Eq.~(\ref{e:contact_neutrino_ex_potential}) and (\ref{e:gen_expval}), 
\begin{equation}\label{e:delE_S1}
    \delta E_{n^3 S_1} = -\frac{G_F^2 M_Z^2 Z^3}{12\pi^3 n^3 \tilde{a}_B^3} \bigg(a_e^2 - \frac{2}{3} b_e^2\bigg).
\end{equation}
The energy shift for muonium $1s-2s$ ($Z=1$ and $\tilde{a}_B = a_B$) evaluates to
\begin{equation}
    \delta E = -2.00\times10^{-16}\ \text{eV}.
\end{equation}
From Ref. \cite{MuoniumMeyer2000}, the maximal difference between experimental and theoretical muonium $1^3 S_1 - 2^3 S_1$ results is $E_{\rm{exp}} - E_{\rm{thr}} = -6.4\times10^{-8}$ eV. Replacing $G_F^2$ with $G_{\rm{eff}}^2$, we find the constraint on the neutrino-mediated potential in muonium $1s-2s$
\begin{equation}
    G_{\rm{eff}}^2 \leq 3.2\times10^8 G_F^2.
\end{equation}
This constraint from muonium spectroscopy is a new result. It may be improved significantly in the near future with new experimental results from the ongoing experiment Mu-MASS \cite{MuMASS}, which aims at improving $1s-2s$ muonium spectroscopy by several orders of magnitude.

We also find that the energy shift for positronium $1s-2s$ ($Z=1$ and $\tilde{a}_B = 2a_B$) is
\begin{equation}
    \delta E = -2.50\times10^{-17}\ \text{eV}.
\end{equation}
The maximal difference between experiment \cite{Ps1s2sExp} and QED calculation \cite{PsThr}  for the positronium $1s-2s$ energy shift is $E_{\rm{exp}} - E_{\rm{thr}} = -3.7\times10^{-8}$ eV. Therefore, we find the constraint on the effective interaction constant
\begin{equation}
    G_{\rm{eff}}^2 \leq 1.5\times10^9 G_F^2.
\end{equation}
Compared to the positronium $1s-2s$ constraint of Ref. \cite{StadnikPRL2018}, our constraint is 6 times weaker due to a more accurate  treatment of the potential at small distances.

\subsection{Muonium and Positronium Ground-State Hyperfine Splitting}
To find constraints from hyperfine splitting (HFS), we calculate the energy shift for muonium and positronium $n^1 S_0$ states using Eq.~(\ref{e:contact_neutrino_ex_potential}) and (\ref{e:gen_expval}),
\begin{equation}\label{e:delE_S0}
    \delta E_{n^3 S_0} = -\frac{G_F^2 M_Z^2 Z^3}{12\pi^3 n^3 \tilde{a}_B^3} \bigg(a_e^2 + 2 b_e^2\bigg).
\end{equation}

Using Eq.~(\ref{e:delE_S1}) and (\ref{e:delE_S0}), we calculate the energy shift for muonium ground-state hyperfine splitting
\begin{equation}
    \delta E = -1.33\times10^{-15}\ \text{eV}.
\end{equation}
The maximal difference between experiment \cite{MuHFSExp} and QED calculation \cite{MuHFSThr1,MuHFSThr2} of muonium ground-state hyperfine splitting is $E_{\rm{exp}} - E_{\rm{thr}} = -1.5\times10^{-12}$ eV. Replacing $G_F^2$ with $G_{\rm{eff}}^2$, we find the constraint
\begin{equation}
    G_{\rm{eff}}^2 \leq 1.1\times10^3 G_F^2.
\end{equation}
We should add that preprint   \cite{arXiv:1810.05429} contains calculation of the electroweak  corrections to the muonium hyperfine structure. Our aim is different: to investigate  effects of the neutrino-exchange potential in atomic systems  and compare corresponding results  with the results of the macroscopic measurements  of this potential.  Our method is certainly not the adequate one for the accurate calculations of all electroweak radiative corrections.

For positronium ground-state hyperfine splitting, we calculate the energy shift to be
\begin{equation}
    \delta E = -1.78\times10^{-16}\ \text{eV}.
\end{equation}
The corresponding maximal difference between experiment \cite{PsHFSExp} and QED calculation \cite{PsThr}  is $E_{\rm{exp}} - E_{\rm{thr}} = -1.6\times10^{-8}$ eV. Therefore, we find the constraint on the effective interaction constant
\begin{equation}
    G_{\rm{eff}}^2 \leq 9.0\times10^7 G_F^2.
\end{equation}

Both hyperfine splitting constraints are 6 times weaker  than those obtained in Ref. \cite{StadnikPRL2018} due to a more accurate treatment of the potential at small distances.

\subsection{Deuteron Binding Energy}
The wave function of the deuteron may be found using the short range nature of the strong interaction and relatively small binding energy of the deuteron.  Outside the interaction range, we use  solution to the Schrödinger equation for zero potential. Within the interaction range $r_0 = 1.2$ fm, the wave function has a constant value for $s$ orbital. Therefore, the wave function is given by
\begin{equation}\label{e:deut_wavefunc}
    \psi(r) = \left\{\begin{array}{ll}
         \frac{B e^{-\kappa r}}{r} & \text{ for } r>r_0,\\
         \frac{B\, J(0)}{r_0} &  \text{ for } r<r_0,
    \end{array}\right.
\end{equation}
where the normalisation constant $B$ is given by $4\pi B^2 = 2\kappa$ for $\kappa = \sqrt{2m |E|} = 4.56\times10^7$ eV (reduced mass $m = m_p/2$ and binding energy $|E| = 2.22$ MeV). The Jastrow factor, $J(0) = 0.4$ \cite{Jastrow}, is included to account for the nucleon repulsion at short distance. Using perturbation theory for a contact potential $g\delta(\mathbf{r})$ 
\begin{equation}
    \langle \psi | g\delta(\mathbf{r}) | \psi \rangle = \frac{g \kappa J(0)^2}{2\pi r_0^2}.
\end{equation}
Substituting $g$ for the neutrino-mediated potential in Eq.~(\ref{e:contact_neutrino_ex_potential}), we obtain the energy shift for the deuteron binding energy
\begin{equation}
    \delta E = -\frac{G_F^2 M_Z^2 \kappa J(0)^2}{24\pi^3 r_0^2} \bigg(a_n a_p - \frac{2}{3} b_n b_p\bigg),
\end{equation}
which evaluates to
\begin{equation}
    \delta E = -1.10\times10^{-3}\ \text{eV}.
\end{equation}

Following Ref. \cite{StadnikPRL2018}, we take difference between experimental \cite{DeuteronExp1999} and theoretical \cite{DeuteronThr2015} results as $E_{\rm{exp}} - E_{\rm{thr}} = -13.7$ eV. This gives 
\begin{equation}
   G_{\rm{eff}}^2 \leq 1.2\times10^4 G_F^2.
\end{equation}
This constraint is 4 orders of magnitude stronger than previously calculated in Ref. \cite{StadnikPRL2018}. This is mainly due to the $Z$ boson propagator cut-off ($Z$ boson Compton wavelength)  instead of the nuclear radius cut-off in Ref. \cite{StadnikPRL2018}. Formally, this looks like the second strongest constraint among two-body systems (the strongest constraint comes from muonium HFS). However, deuteron is  a system with the strong interaction and this constraint is probably less reliable than the  constraints from the lepton systems.

\begin{table}[tb]
\begin{tabular}{c c}
    \hline
    Case & \, $G_{\rm{eff}}^2/G_F^2$ \, \\\hline
    Hydrogen $1s-2s$ & $6.1\times10^4$ \\
    Hydrogen $1s-3s$ & $5.5\times10^4$ \\
    Muonium $1s-2s$ & $3.2\times10^8$ \\
    Positronium $1s-2s$ & $1.5\times10^9$ \\
    Muonium HFS & $1.1\times10^3$ \\
    Positronium HFS & $9.0\times10^7$ \\
    Deuteron Binding Energy & $1.2\times10^4$ \\\hline
\end{tabular}
\caption{Summary of constraints $G_{\rm{eff}}^2/G_F^2$ on neutrino-mediated potential (see Eq.~(\ref{e:contact_neutrino_ex_potential})) in simple systems.}
\label{tab:SimpleSystemResults}\end{table}

\section{Energy Shift in Molecular Hydrogen Systems}\label{SectionMolecules}
We also examine the constraints obtained from molecular systems for the neutrino-exchange interaction. On the molecular scale, it is sufficient to use Eq.~(\ref{e:neutrino_ex_potential}) (only $a_1 a_2$ contribute) as the nuclei are separated by a distance at least $a_B$. Additionally, this also means that only neutrinos contribute to the interaction, so we use $N_{\rm{eff}} = 3$. We consider molecular hydrogen systems and thus present the potential
\begin{eqnarray}\label{e:molecule_neutrino_ex_potential}
    V_\nu^M (r) &=& \frac{G_F^2 N_{\rm{eff}} a_1^{(1)} a_2^{(2)}}{4 \pi^3 r^5}   \equiv \frac{g}{r^5},
\end{eqnarray}
where the interacting particles are nucleons. Ref. \cite{Salumbides2015} used precision molecular spectroscopy to obtain constraints with regards to gravity in extra dimensions, including a potential with $1/r^5$ dependence. This was done similarly to our method in Section \ref{SectionSimpleSystems}, where the difference between theoretical and experimental results  was used to obtain  constraints on the interaction strength.

We utilise the findings of Ref. \cite{Salumbides2015} to obtain constraints from the systems H$_2$, D$_2$, and HD$^+$. 
Values for the chosen systems are $a_p^{(1)} a_p^{(2)} = 1.5\times10^{-3}$, $a_n^{(1)} a_n^{(2)} = 0.75$, and $a_p^{(1)} a_n^{(2)} = -3.3\times10^{-2}$.
Using the results of Ref. \cite{Salumbides2015}, we find constraints on the interaction strength of the neutrino-exchange interaction presented in Table \ref{tab:MoleculeResults}.

\begin{table}[tb]
\begin{tabular}{c c c c}
    \hline
    Case            & $g$ (GeV$^{-4}$)      & \, $G_{\rm{eff}}^2/G_F^2$ \,  & Ref.\\\hline
    H$_2$ (1-0)     & $  6\times10^{12}$   & $4\times10^{27}$            & \cite{H2_10_a,H2_10_b}\\
    H$_2$ ($D_0$)   & $2\times10^{12}$   & $1.3\times10^{27}$            & \cite{H2_D0}\\
    D$_2$ ($D_0$)   & $ 2\times10^{12}$   & $2.5\times10^{24}$            & \cite{D2_D0}\\
    HD$^+$ (4-0)    & $ 2\times10^{12}$   & $5  \times10^{25}$            & \cite{HD_40}\\\hline
\end{tabular}
\caption{ Summary of constraints $G_{\rm{eff}}^2/G_F^2$ on neutrino-mediated potential 
(see Eq.~(\ref{e:molecule_neutrino_ex_potential})) in molecular systems. Values $(i \to j)$ correspond to vibrational transitions, $D_0$ corresponds to the dissociation limit. References correspond to experimental and theoretical works used to determine difference between  experimental and theoretical results for the transition energies.}
\label{tab:MoleculeResults}\end{table}

\section{The Yukawa Potential}
In addition to constraints on the neutrino-exchange interaction, one can also use spectroscopy to place constraints on a fifth force from beyond the Standard Model. The force can be phenomenologically parameterised by a Yukawa-type potential
\begin{equation}
    V_5 = \frac{\beta e^{-mcr/\hbar}}{r},
\end{equation}
where $\beta$ is the coupling strength and $m$ is the mass of a scalar particle mediating Yukawa-type interaction.  Constraints were obtained  using molecular hydrogen spectroscopy in Ref. \cite{Yukawa_Molecule} and atomic hydrogen $1s-2s$ spectroscopy in Ref. \cite{Yukawa_H}. It was found that atomic systems provide stronger limits than molecular systems. We improve upon Ref. \cite{Yukawa_H} by using more recent data and provide additional constraints from muonium $1s-2s$, positronium $1s-2s$, hydrogen $1s-3s$, and deuteron.

\begin{figure}[htb]
    \includegraphics[width=0.45\textwidth]{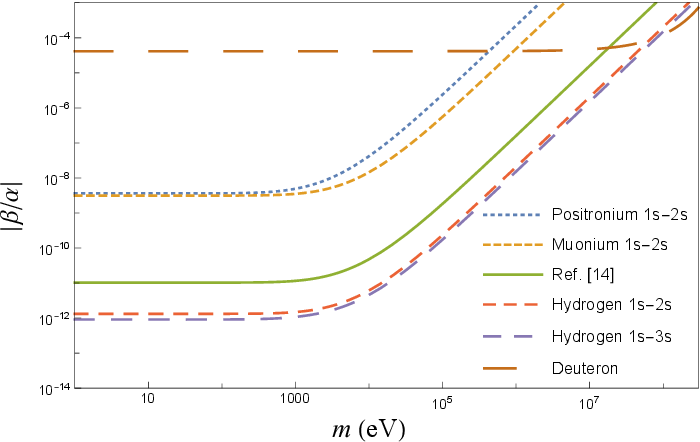}
    \caption{\label{fig:yukawa} Upper limits from atomic spectroscopy on the coupling strength of a Yukawa-type potential $\beta$ relative to the fine structure constant $\alpha$ as a function of the force-mediating scalar particle mass $m$.}
\end{figure}

The energy shifts for $1s-2s$ and $1s-3s$ transitions found from the expectation values of the Yukawa potential for hydrogen-like atoms (in units $\hbar = c = 1$)
\begin{equation}
    \delta E_{1s-2s} = \frac{\beta}{4\tilde{a}_B} \left[ \frac{1 + 2 \tilde{a}_B^2 m^2}{(1 + \tilde{a}_B m)^4} - \frac{16}{(2 + \tilde{a}_B m)^2} \right],
\end{equation}
\begin{eqnarray}
    \delta E_{1s-3s} =&& \frac{4\beta}{9\tilde{a}_B} \left[ \frac{16 + 27 \tilde{a}_B^2 m^2 (8 + 9 \tilde{a}_B^2 m^2)}{(2 + 3 \tilde{a}_B m)^6} \right. \nonumber\\
    &&\left. - \frac{9}{(2 + \tilde{a}_B m)^2} \right],
\end{eqnarray}
where $\tilde{a}_B$ is the reduced Bohr radius. Note that in both transitions, in the large mass limit we observe $\delta E \propto \beta/m^2$. The energy shift for deuteron, using the wave function in Eq. \ref{e:deut_wavefunc}, is given by
\begin{equation}
    \delta E = - 2 \beta \kappa {\rm Ei}\big(-(m+2\kappa)r_0\big),
\end{equation}
where we have the exponential integral ${\rm Ei}(x) = -\int_{-x}^\infty e^{-t}/t \ dt$. To obtain constraints we use limits on energy shifts for hydrogen, muonium, and positronium transitions and deuteron binding energy presented in Section \ref{SectionSimpleSystems}.
For comparison, we include the previous constraint from Ref. \cite{Yukawa_H} which used $|\delta E| = 2.1\times10^{-10}$ eV. Our results are presented in Figure \ref{fig:yukawa}. Hydrogen $1s-3s$ gives the strongest constraints at small masses $|\beta/\alpha| < 9.2\times10^{-13}$, while at large masses this constraint is $|\beta/\alpha| < 1.5\times10^{-20}\ (m/\rm{eV})^2$.

\section{Conclusion}\label{SectionConclusion}
Our work is motivated by the Stadnik paper \cite{StadnikPRL2018}, which demonstrated that the sensitivity of atomic spectral data to the neutrino mediated potential, introduced in Refs. \cite{GamowPR1937,Feynman,FeinbergPR1968,FeinbergPR1989,HsuPRD1994}, is up to 18 orders of magnitude better than the sensitivity of macroscopic experiments to this potential. However, an oversimplified  cut-off treatment of this potential at small distance has  led to inaccurate results in Ref. \cite{StadnikPRL2018}, especially in systems with finite size of the particles. For example, cut-off at the nuclear  radius gave limits on the interaction strength  which are five orders of magnitude weaker than the limits obtained in the present work. We argue that one should firstly build  effective interaction between  point-like particles, like electrons and quarks. On the second step this interaction should be integrated over the nuclear volume. 

In this paper, we calculated energy shifts, produced by neutrino potentials, and extracted limits on the strength of this potential from hydrogen-like systems, namely hydrogen, muonium, positronium, and deuteron. We also extracted constraints from spectra of H$_2$, D$_2$ and HD$^+$ molecules. Following Ref. \cite{StadnikPRL2018}, we presented our results as constraints on the ratio of the effective strength of the neutrino-mediated potential $G_{\rm{eff}}^2$ to the squared Fermi constant $G_F^2$. The best limit was obtained from the muonium hyperfine structure, $G_{\rm{eff}}^2/G_F^2 < 10^3$. Our constraints are expected to be significantly enhanced in the near future, for example, with the muonium $1s-2s$ measurement predicted to be improved by three orders of magnitude by the currently ongoing experiment MuMASS \cite{MuMASS}. Constraints from atomic spectroscopy on the coupling strength and mass of a new scalar particle mediating a fifth force are also obtained and are an order of magnitude stronger compared to Ref. \cite{Yukawa_H}.

\section*{Acknowledgements}
 This work was supported by the Australian Research Council Grants No. DP190100974 and DP200100150.


\begin{thebibliography}{46}
\expandafter\ifx\csname natexlab\endcsname\relax\def\natexlab#1{#1}\fi
\expandafter\ifx\csname bibnamefont\endcsname\relax
  \def\bibnamefont#1{#1}\fi
\expandafter\ifx\csname bibfnamefont\endcsname\relax
  \def\bibfnamefont#1{#1}\fi
\expandafter\ifx\csname citenamefont\endcsname\relax
  \def\citenamefont#1{#1}\fi
\expandafter\ifx\csname url\endcsname\relax
  \def\url#1{\texttt{#1}}\fi
\expandafter\ifx\csname urlprefix\endcsname\relax\def\urlprefix{URL }\fi
\providecommand{\bibinfo}[2]{#2}
\providecommand{\eprint}[2][]{\url{#2}}

\bibitem[{\citenamefont{Gamow and Teller}(1937)}]{GamowPR1937}
\bibinfo{author}{\bibfnamefont{G.}~\bibnamefont{Gamow}} \bibnamefont{and}
  \bibinfo{author}{\bibfnamefont{E.}~\bibnamefont{Teller}},
  \bibinfo{journal}{Phys. Rev.} \textbf{\bibinfo{volume}{51}},
  \bibinfo{pages}{289} (\bibinfo{year}{1937}),
  \urlprefix\url{https://link.aps.org/doi/10.1103/PhysRev.51.289}.

\bibitem[{\citenamefont{Feynman}(1996)}]{Feynman}
\bibinfo{author}{\bibfnamefont{R.~P.} \bibnamefont{Feynman}},
  \emph{\bibinfo{title}{Feynman Lectures on Gravitation}}
  (\bibinfo{publisher}{Addison-Wesley}, \bibinfo{address}{Reading, MA},
  \bibinfo{year}{1996}).

\bibitem[{\citenamefont{Feinberg and Sucher}(1968)}]{FeinbergPR1968}
\bibinfo{author}{\bibfnamefont{G.}~\bibnamefont{Feinberg}} \bibnamefont{and}
  \bibinfo{author}{\bibfnamefont{J.}~\bibnamefont{Sucher}},
  \bibinfo{journal}{Phys. Rev., 166: 1638-44(Feb. 25, 1968).}
  (\bibinfo{year}{1968}), \urlprefix\url{https://www.osti.gov/biblio/4554464}.

\bibitem[{\citenamefont{Feinberg et~al.}(1989)\citenamefont{Feinberg, Sucher,
  and Au}}]{FeinbergPR1989}
\bibinfo{author}{\bibfnamefont{G.}~\bibnamefont{Feinberg}},
  \bibinfo{author}{\bibfnamefont{J.}~\bibnamefont{Sucher}}, \bibnamefont{and}
  \bibinfo{author}{\bibfnamefont{C.-K.} \bibnamefont{Au}},
  \bibinfo{journal}{Physics Reports} \textbf{\bibinfo{volume}{180}},
  \bibinfo{pages}{83} (\bibinfo{year}{1989}), ISSN \bibinfo{issn}{0370-1573},
  \urlprefix\url{https://www.sciencedirect.com/science/article/pii/0370157389901117}.

\bibitem[{\citenamefont{Hsu and Sikivie}(1994)}]{HsuPRD1994}
\bibinfo{author}{\bibfnamefont{S.~D.~H.} \bibnamefont{Hsu}} \bibnamefont{and}
  \bibinfo{author}{\bibfnamefont{P.}~\bibnamefont{Sikivie}},
  \bibinfo{journal}{Physical Review D} \textbf{\bibinfo{volume}{49}},
  \bibinfo{pages}{4951} (\bibinfo{year}{1994}), ISSN \bibinfo{issn}{0556-2821},
  \urlprefix\url{http://dx.doi.org/10.1103/PhysRevD.49.4951}.

\bibitem[{\citenamefont{Kapner et~al.}(2007)\citenamefont{Kapner, Cook,
  Adelberger, Gundlach, Heckel, Hoyle, and Swanson}}]{Kapner2007}
\bibinfo{author}{\bibfnamefont{D.~J.} \bibnamefont{Kapner}},
  \bibinfo{author}{\bibfnamefont{T.~S.} \bibnamefont{Cook}},
  \bibinfo{author}{\bibfnamefont{E.~G.} \bibnamefont{Adelberger}},
  \bibinfo{author}{\bibfnamefont{J.~H.} \bibnamefont{Gundlach}},
  \bibinfo{author}{\bibfnamefont{B.~R.} \bibnamefont{Heckel}},
  \bibinfo{author}{\bibfnamefont{C.~D.} \bibnamefont{Hoyle}}, \bibnamefont{and}
  \bibinfo{author}{\bibfnamefont{H.~E.} \bibnamefont{Swanson}},
  \bibinfo{journal}{Phys. Rev. Lett.} \textbf{\bibinfo{volume}{98}},
  \bibinfo{pages}{021101} (\bibinfo{year}{2007}).

\bibitem[{\citenamefont{Adelberger et~al.}(2007)\citenamefont{Adelberger,
  Heckel, Hoedl, Hoyle, Kapner, and Upadhye}}]{Adelberger2007}
\bibinfo{author}{\bibfnamefont{E.~G.} \bibnamefont{Adelberger}},
  \bibinfo{author}{\bibfnamefont{B.~R.} \bibnamefont{Heckel}},
  \bibinfo{author}{\bibfnamefont{S.}~\bibnamefont{Hoedl}},
  \bibinfo{author}{\bibfnamefont{C.~D.} \bibnamefont{Hoyle}},
  \bibinfo{author}{\bibfnamefont{D.~J.} \bibnamefont{Kapner}},
  \bibnamefont{and} \bibinfo{author}{\bibfnamefont{A.}~\bibnamefont{Upadhye}},
  \bibinfo{journal}{Phys. Rev. Lett.} \textbf{\bibinfo{volume}{98}},
  \bibinfo{pages}{131104} (\bibinfo{year}{2007}).

\bibitem[{\citenamefont{Chen et~al.}(2016)\citenamefont{Chen, Tham, Krause,
  Lopez, Fischbach, and Decca}}]{Chen2016}
\bibinfo{author}{\bibfnamefont{Y.-J.} \bibnamefont{Chen}},
  \bibinfo{author}{\bibfnamefont{W.~K.} \bibnamefont{Tham}},
  \bibinfo{author}{\bibfnamefont{D.~E.} \bibnamefont{Krause}},
  \bibinfo{author}{\bibfnamefont{D.}~\bibnamefont{Lopez}},
  \bibinfo{author}{\bibfnamefont{E.}~\bibnamefont{Fischbach}},
  \bibnamefont{and} \bibinfo{author}{\bibfnamefont{R.~S.} \bibnamefont{Decca}},
  \bibinfo{journal}{Phys. Rev. Lett.} \textbf{\bibinfo{volume}{116}},
  \bibinfo{pages}{221102} (\bibinfo{year}{2016}).

\bibitem[{\citenamefont{Vasilakis et~al.}(2009)\citenamefont{Vasilakis, Brown,
  Kornack, and Romalis}}]{Vasilakis2009}
\bibinfo{author}{\bibfnamefont{G.}~\bibnamefont{Vasilakis}},
  \bibinfo{author}{\bibfnamefont{J.~M.} \bibnamefont{Brown}},
  \bibinfo{author}{\bibfnamefont{T.~W.} \bibnamefont{Kornack}},
  \bibnamefont{and} \bibinfo{author}{\bibfnamefont{M.~V.}
  \bibnamefont{Romalis}}, \bibinfo{journal}{Phys. Rev. Lett.}
  \textbf{\bibinfo{volume}{103}}, \bibinfo{pages}{261801}
  (\bibinfo{year}{2009}).

\bibitem[{\citenamefont{Terrano et~al.}(2015)\citenamefont{Terrano, Adelberger,
  Lee, and Heckel}}]{Terrano2015}
\bibinfo{author}{\bibfnamefont{W.~A.} \bibnamefont{Terrano}},
  \bibinfo{author}{\bibfnamefont{E.~G.} \bibnamefont{Adelberger}},
  \bibinfo{author}{\bibfnamefont{J.~G.} \bibnamefont{Lee}}, \bibnamefont{and}
  \bibinfo{author}{\bibfnamefont{B.~R.} \bibnamefont{Heckel}},
  \bibinfo{journal}{Phys. Rev. Lett.} \textbf{\bibinfo{volume}{115}},
  \bibinfo{pages}{201801} (\bibinfo{year}{2015}).

\bibitem[{\citenamefont{Stadnik}(2018)}]{StadnikPRL2018}
\bibinfo{author}{\bibfnamefont{Y.~V.} \bibnamefont{Stadnik}},
  \bibinfo{journal}{Phys. Rev. Lett.} \textbf{\bibinfo{volume}{120}},
  \bibinfo{pages}{223202} (\bibinfo{year}{2018}),
  \urlprefix\url{https://link.aps.org/doi/10.1103/PhysRevLett.120.223202}.

\bibitem[{\citenamefont{Meyer et~al.}(2000)\citenamefont{Meyer, Bagayev, Baird,
  Bakule, Boshier, Breitrück, Cornish, Dychkov, Eaton, Grossmann
  et~al.}}]{MuoniumMeyer2000}
\bibinfo{author}{\bibfnamefont{V.}~\bibnamefont{Meyer}},
  \bibinfo{author}{\bibfnamefont{S.~N.} \bibnamefont{Bagayev}},
  \bibinfo{author}{\bibfnamefont{P.~E.~G.} \bibnamefont{Baird}},
  \bibinfo{author}{\bibfnamefont{P.}~\bibnamefont{Bakule}},
  \bibinfo{author}{\bibfnamefont{M.~G.} \bibnamefont{Boshier}},
  \bibinfo{author}{\bibfnamefont{A.}~\bibnamefont{Breitrück}},
  \bibinfo{author}{\bibfnamefont{S.~L.} \bibnamefont{Cornish}},
  \bibinfo{author}{\bibfnamefont{S.}~\bibnamefont{Dychkov}},
  \bibinfo{author}{\bibfnamefont{G.~H.} \bibnamefont{Eaton}},
  \bibinfo{author}{\bibfnamefont{A.}~\bibnamefont{Grossmann}},
  \bibnamefont{et~al.}, \bibinfo{journal}{Physical Review Letters}
  \textbf{\bibinfo{volume}{84}}, \bibinfo{pages}{1136} (\bibinfo{year}{2000}),
  ISSN \bibinfo{issn}{1079-7114},
  \urlprefix\url{http://dx.doi.org/10.1103/PhysRevLett.84.1136}.

\bibitem[{\citenamefont{Karshenboim}(2005)}]{PositroniumKarshenboim2000}
\bibinfo{author}{\bibfnamefont{S.~G.} \bibnamefont{Karshenboim}},
  \bibinfo{journal}{Physics Reports} \textbf{\bibinfo{volume}{422}},
  \bibinfo{pages}{1} (\bibinfo{year}{2005}), ISSN \bibinfo{issn}{0370-1573},
  \urlprefix\url{https://www.sciencedirect.com/science/article/pii/S0370157305003637}.

\bibitem[{\citenamefont{Ubachs et~al.}(2013)\citenamefont{Ubachs, Vassen,
  Salumbides, and Eikema}}]{Yukawa_H}
\bibinfo{author}{\bibfnamefont{W.}~\bibnamefont{Ubachs}},
  \bibinfo{author}{\bibfnamefont{W.}~\bibnamefont{Vassen}},
  \bibinfo{author}{\bibfnamefont{E.~J.}~\bibnamefont{Salumbides}}, \bibnamefont{and}
  \bibinfo{author}{\bibfnamefont{K.~S.~E.}~\bibnamefont{Eikema}},
  \bibinfo{journal}{Ann. Phys. (Berlin)} \textbf{\bibinfo{volume}{525}},
  \bibinfo{pages}{A113–A115} (\bibinfo{year}{2013}),
  \urlprefix\url{https://doi.org/10.1002/andp.201300730}.

\bibitem[{\citenamefont{Xu and Yu}(2022)}]{Xu2022}
\bibinfo{author}{\bibfnamefont{X.-J.} \bibnamefont{Xu}} \bibnamefont{and}
  \bibinfo{author}{\bibfnamefont{B.}~\bibnamefont{Yu}},
  \bibinfo{journal}{Journal of High Energy Physics}
  \textbf{\bibinfo{volume}{2022}} (\bibinfo{year}{2022}),
  \urlprefix\url{https://doi.org/10.48550/arXiv.2112.03060}.

\bibitem[{\citenamefont{Grifols et~al.}(1996)\citenamefont{Grifols, Masso, and
  Toldra}}]{Grifols}
\bibinfo{author}{\bibfnamefont{J.~A.} \bibnamefont{Grifols}},
  \bibinfo{author}{\bibfnamefont{E.}~\bibnamefont{Masso}}, \bibnamefont{and}
  \bibinfo{author}{\bibfnamefont{R.}~\bibnamefont{Toldra}},
  \bibinfo{journal}{Phys. Lett. B} \textbf{\bibinfo{volume}{389}},
  \bibinfo{pages}{563} (\bibinfo{year}{1996}).

\bibitem[{\citenamefont{Tanabashi et~al.}(2018)\citenamefont{Tanabashi,
  Hagiwara, Hikasa, Nakamura, Sumino, Takahashi, Tanaka, Agashe, Aielli, Amsler
  et~al.}}]{SM}
\bibinfo{author}{\bibfnamefont{M.}~\bibnamefont{Tanabashi}},
  \bibinfo{author}{\bibfnamefont{K.}~\bibnamefont{Hagiwara}},
  \bibinfo{author}{\bibfnamefont{K.}~\bibnamefont{Hikasa}},
  \bibinfo{author}{\bibfnamefont{K.}~\bibnamefont{Nakamura}},
  \bibinfo{author}{\bibfnamefont{Y.}~\bibnamefont{Sumino}},
  \bibinfo{author}{\bibfnamefont{F.}~\bibnamefont{Takahashi}},
  \bibinfo{author}{\bibfnamefont{J.}~\bibnamefont{Tanaka}},
  \bibinfo{author}{\bibfnamefont{K.}~\bibnamefont{Agashe}},
  \bibinfo{author}{\bibfnamefont{G.}~\bibnamefont{Aielli}},
  \bibinfo{author}{\bibfnamefont{C.}~\bibnamefont{Amsler}},
  \bibnamefont{et~al.} (\bibinfo{collaboration}{Particle Data Group}),
  \bibinfo{journal}{Phys. Rev. D} \textbf{\bibinfo{volume}{98}},
  \bibinfo{pages}{030001} (\bibinfo{year}{2018}),
  \urlprefix\url{https://link.aps.org/doi/10.1103/PhysRevD.98.030001}.


\bibitem[{\citenamefont{Ahmadi et~al.}(2018)\citenamefont{Ahmadi, Alves,
  Baker, and et~al.}}]{H1s2s}
\bibinfo{author}{\bibfnamefont{W.}~\bibnamefont{Ahmadi}},
  \bibinfo{author}{\bibfnamefont{W.}~\bibnamefont{Alves}},
  \bibinfo{author}{\bibfnamefont{E.~J.}~\bibnamefont{Baker}},
  \bibnamefont{et~al.},
  \bibinfo{journal}{Nature} \textbf{\bibinfo{volume}{557}},
  \bibinfo{pages}{71-75} (\bibinfo{year}{2018}),
  \urlprefix\url{https://doi.org/10.1038/s41586-018-0017-2}.
  
  \bibitem[{\citenamefont{Fleurbaey et~al.}(2018)\citenamefont{Fleurbaey, Galtier,
  Thomas, Bonnaud, Julien, Biraben, Nez, Abgrall and Guéna}}]{H1s3s}
\bibinfo{author}{\bibfnamefont{H.}~\bibnamefont{Fleurbaey}},
  \bibinfo{author}{\bibfnamefont{S.}~\bibnamefont{Galtier}},
  \bibinfo{author}{\bibfnamefont{S.}~\bibnamefont{Thomas}},
  \bibinfo{author}{\bibfnamefont{M.}~\bibnamefont{Bonnaud}},
  \bibinfo{author}{\bibfnamefont{L.}~\bibnamefont{Julien}},
  \bibinfo{author}{\bibfnamefont{F.}~\bibnamefont{Biraben}},
  \bibinfo{author}{\bibfnamefont{F.}~\bibnamefont{Nez}},
  \bibinfo{author}{\bibfnamefont{M.}~\bibnamefont{Abgrall}}, \bibnamefont{and}
  \bibinfo{author}{\bibfnamefont{J.}~\bibnamefont{Guéna}},
  \bibinfo{journal}{Phys. Rev. Lett.} \textbf{\bibinfo{volume}{120}},
  \bibinfo{pages}{183001 } (\bibinfo{year}{2018}),
  \urlprefix\url{https://doi.org/10.1103/PhysRevLett.120.183001}.

\bibitem[{\citenamefont{Crivelli}(2018)}]{MuMASS}
\bibinfo{author}{\bibfnamefont{P.}~\bibnamefont{Crivelli}},
  \bibinfo{journal}{Hyperfine Interactions} \textbf{\bibinfo{volume}{239}}
  (\bibinfo{year}{2018}), ISSN \bibinfo{issn}{1572-9540},
  \urlprefix\url{http://dx.doi.org/10.1007/s10751-018-1525-z}.

\bibitem[{\citenamefont{Fee et~al.}(1993)\citenamefont{Fee, Mills, Chu, Shaw,
  Danzmann, Chichester, and Zuckerman}}]{Ps1s2sExp}
\bibinfo{author}{\bibfnamefont{M.~S.} \bibnamefont{Fee}},
  \bibinfo{author}{\bibfnamefont{A.~P.} \bibnamefont{Mills}},
  \bibinfo{author}{\bibfnamefont{S.}~\bibnamefont{Chu}},
  \bibinfo{author}{\bibfnamefont{E.~D.} \bibnamefont{Shaw}},
  \bibinfo{author}{\bibfnamefont{K.}~\bibnamefont{Danzmann}},
  \bibinfo{author}{\bibfnamefont{R.~J.} \bibnamefont{Chichester}},
  \bibnamefont{and} \bibinfo{author}{\bibfnamefont{D.~M.}
  \bibnamefont{Zuckerman}}, \bibinfo{journal}{Phys. Rev. Lett.}
  \textbf{\bibinfo{volume}{70}}, \bibinfo{pages}{1397} (\bibinfo{year}{1993}),
  \urlprefix\url{https://link.aps.org/doi/10.1103/PhysRevLett.70.1397}.

\bibitem[{\citenamefont{Czarnecki et~al.}(1999)\citenamefont{Czarnecki,
  Melnikov, and Yelkhovsky}}]{PsThr}
\bibinfo{author}{\bibfnamefont{A.}~\bibnamefont{Czarnecki}},
  \bibinfo{author}{\bibfnamefont{K.}~\bibnamefont{Melnikov}}, \bibnamefont{and}
  \bibinfo{author}{\bibfnamefont{A.}~\bibnamefont{Yelkhovsky}},
  \bibinfo{journal}{Phys. Rev. A} \textbf{\bibinfo{volume}{59}},
  \bibinfo{pages}{4316} (\bibinfo{year}{1999}),
  \urlprefix\url{https://link.aps.org/doi/10.1103/PhysRevA.59.4316}.

\bibitem[{\citenamefont{Liu et~al.}(1999)\citenamefont{Liu, Boshier, Dhawan,
  van Dyck, Egan, Fei, Grosse~Perdekamp, Hughes, Janousch, Jungmann
  et~al.}}]{MuHFSExp}
\bibinfo{author}{\bibfnamefont{W.}~\bibnamefont{Liu}},
  \bibinfo{author}{\bibfnamefont{M.~G.} \bibnamefont{Boshier}},
  \bibinfo{author}{\bibfnamefont{S.}~\bibnamefont{Dhawan}},
  \bibinfo{author}{\bibfnamefont{O.}~\bibnamefont{van Dyck}},
  \bibinfo{author}{\bibfnamefont{P.}~\bibnamefont{Egan}},
  \bibinfo{author}{\bibfnamefont{X.}~\bibnamefont{Fei}},
  \bibinfo{author}{\bibfnamefont{M.}~\bibnamefont{Grosse~Perdekamp}},
  \bibinfo{author}{\bibfnamefont{V.~W.} \bibnamefont{Hughes}},
  \bibinfo{author}{\bibfnamefont{M.}~\bibnamefont{Janousch}},
  \bibinfo{author}{\bibfnamefont{K.}~\bibnamefont{Jungmann}},
  \bibnamefont{et~al.}, \bibinfo{journal}{Phys. Rev. Lett.}
  \textbf{\bibinfo{volume}{82}}, \bibinfo{pages}{711} (\bibinfo{year}{1999}),
  \urlprefix\url{https://link.aps.org/doi/10.1103/PhysRevLett.82.711}.

\bibitem[{\citenamefont{Czarnecki et~al.}(2002)\citenamefont{Czarnecki,
  Eidelman, and Karshenboim}}]{MuHFSThr1}
\bibinfo{author}{\bibfnamefont{A.}~\bibnamefont{Czarnecki}},
  \bibinfo{author}{\bibfnamefont{S.~I.} \bibnamefont{Eidelman}},
  \bibnamefont{and} \bibinfo{author}{\bibfnamefont{S.~G.}
  \bibnamefont{Karshenboim}}, \bibinfo{journal}{Phys. Rev. D}
  \textbf{\bibinfo{volume}{65}}, \bibinfo{pages}{053004}
  (\bibinfo{year}{2002}),
  \urlprefix\url{https://link.aps.org/doi/10.1103/PhysRevD.65.053004}.

\bibitem[{\citenamefont{Mohr et~al.}(2016)\citenamefont{Mohr, Newell, and
  Taylor}}]{MuHFSThr2}
\bibinfo{author}{\bibfnamefont{P.~J.} \bibnamefont{Mohr}},
  \bibinfo{author}{\bibfnamefont{D.~B.} \bibnamefont{Newell}},
  \bibnamefont{and} \bibinfo{author}{\bibfnamefont{B.~N.}
  \bibnamefont{Taylor}}, \bibinfo{journal}{Rev. Mod. Phys.}
  \textbf{\bibinfo{volume}{88}}, \bibinfo{pages}{035009}
  (\bibinfo{year}{2016}),
  \urlprefix\url{https://link.aps.org/doi/10.1103/RevModPhys.88.035009}.

\bibitem[{\citenamefont{Asaka et~al.}(2018)\citenamefont{Asaka, Tanaka,
  Tsumura, and M.Yoshimura}}]{arXiv:1810.05429}
\bibinfo{author}{\bibfnamefont{T.}~\bibnamefont{Asaka}},
  \bibinfo{author}{\bibfnamefont{M.}~\bibnamefont{Tanaka}},
  \bibinfo{author}{\bibfnamefont{K.}~\bibnamefont{Tsumura}}, \bibnamefont{and}
  \bibinfo{author}{\bibnamefont{M.Yoshimura}}, \emph{\bibinfo{title}{Precision
  electroweak shift of muonium hyperfine splitting}} (\bibinfo{year}{2018}),
  \urlprefix\url{https://arxiv.org/abs/1810.05425}.

\bibitem[{\citenamefont{Ritter et~al.}(1984)\citenamefont{Ritter, Egan, Hughes,
  and Woodle}}]{PsHFSExp}
\bibinfo{author}{\bibfnamefont{M.~W.} \bibnamefont{Ritter}},
  \bibinfo{author}{\bibfnamefont{P.~O.} \bibnamefont{Egan}},
  \bibinfo{author}{\bibfnamefont{V.~W.} \bibnamefont{Hughes}},
  \bibnamefont{and} \bibinfo{author}{\bibfnamefont{K.~A.}
  \bibnamefont{Woodle}}, \bibinfo{journal}{Phys. Rev. A}
  \textbf{\bibinfo{volume}{30}}, \bibinfo{pages}{1331} (\bibinfo{year}{1984}),
  \urlprefix\url{https://link.aps.org/doi/10.1103/PhysRevA.30.1331}.

\bibitem[{\citenamefont{Dmitriev et~al.}(1983)\citenamefont{Dmitriev, Flambaum,
  Sushkov, and Telitsin}}]{Jastrow}
\bibinfo{author}{\bibfnamefont{V.~F.} \bibnamefont{Dmitriev}},
  \bibinfo{author}{\bibfnamefont{V.~V.} \bibnamefont{Flambaum}},
  \bibinfo{author}{\bibfnamefont{O.~P.} \bibnamefont{Sushkov}},
  \bibnamefont{and} \bibinfo{author}{\bibfnamefont{V.~B.}
  \bibnamefont{Telitsin}}, \bibinfo{journal}{Physics Letters B}
  \textbf{\bibinfo{volume}{125}}, \bibinfo{pages}{1} (\bibinfo{year}{1983}).

\bibitem[{\citenamefont{{Kessler, Jr} et~al.}(1999)\citenamefont{{Kessler, Jr},
  Dewey, Deslattes, Henins, Börner, Jentschel, Doll, and
  Lehmann}}]{DeuteronExp1999}
\bibinfo{author}{\bibfnamefont{E.}~\bibnamefont{{Kessler, Jr}}},
  \bibinfo{author}{\bibfnamefont{M.}~\bibnamefont{Dewey}},
  \bibinfo{author}{\bibfnamefont{R.}~\bibnamefont{Deslattes}},
  \bibinfo{author}{\bibfnamefont{A.}~\bibnamefont{Henins}},
  \bibinfo{author}{\bibfnamefont{H.}~\bibnamefont{Börner}},
  \bibinfo{author}{\bibfnamefont{M.}~\bibnamefont{Jentschel}},
  \bibinfo{author}{\bibfnamefont{C.}~\bibnamefont{Doll}}, \bibnamefont{and}
  \bibinfo{author}{\bibfnamefont{H.}~\bibnamefont{Lehmann}},
  \bibinfo{journal}{Physics Letters A} \textbf{\bibinfo{volume}{255}},
  \bibinfo{pages}{221} (\bibinfo{year}{1999}), ISSN \bibinfo{issn}{0375-9601},
  \urlprefix\url{https://www.sciencedirect.com/science/article/pii/S037596019900078X}.

\bibitem[{\citenamefont{Ekstr\"om et~al.}(2015)\citenamefont{Ekstr\"om, Jansen,
  Wendt, Hagen, Papenbrock, Carlsson, Forss\'en, Hjorth-Jensen, Navr\'atil, and
  Nazarewicz}}]{DeuteronThr2015}
\bibinfo{author}{\bibfnamefont{A.}~\bibnamefont{Ekstr\"om}},
  \bibinfo{author}{\bibfnamefont{G.~R.} \bibnamefont{Jansen}},
  \bibinfo{author}{\bibfnamefont{K.~A.} \bibnamefont{Wendt}},
  \bibinfo{author}{\bibfnamefont{G.}~\bibnamefont{Hagen}},
  \bibinfo{author}{\bibfnamefont{T.}~\bibnamefont{Papenbrock}},
  \bibinfo{author}{\bibfnamefont{B.~D.} \bibnamefont{Carlsson}},
  \bibinfo{author}{\bibfnamefont{C.}~\bibnamefont{Forss\'en}},
  \bibinfo{author}{\bibfnamefont{M.}~\bibnamefont{Hjorth-Jensen}},
  \bibinfo{author}{\bibfnamefont{P.}~\bibnamefont{Navr\'atil}},
  \bibnamefont{and}
  \bibinfo{author}{\bibfnamefont{W.}~\bibnamefont{Nazarewicz}},
  \bibinfo{journal}{Phys. Rev. C} \textbf{\bibinfo{volume}{91}},
  \bibinfo{pages}{051301(R)} (\bibinfo{year}{2015}),
  \urlprefix\url{https://link.aps.org/doi/10.1103/PhysRevC.91.051301}.

\bibitem[{\citenamefont{Salumbides et~al.}(2015)\citenamefont{Salumbides,
  Schellekens, Gato-Rivera, and Ubachs}}]{Salumbides2015}
\bibinfo{author}{\bibfnamefont{E.~J.} \bibnamefont{Salumbides}},
  \bibinfo{author}{\bibfnamefont{A.~N.} \bibnamefont{Schellekens}},
  \bibinfo{author}{\bibfnamefont{B.}~\bibnamefont{Gato-Rivera}},
  \bibnamefont{and} \bibinfo{author}{\bibfnamefont{W.}~\bibnamefont{Ubachs}},
  \bibinfo{journal}{New Journal of Physics} \textbf{\bibinfo{volume}{17}},
  \bibinfo{pages}{033015} (\bibinfo{year}{2015}),
  \urlprefix\url{https://doi.org/10.1088/1367-2630/17/3/033015}.

\bibitem[{\citenamefont{Niu et~al.}(2014)\citenamefont{Niu, Salumbides,
  Dickenson, Eikema, and Ubachs}}]{H2_10_a}
\bibinfo{author}{\bibfnamefont{M.}~\bibnamefont{Niu}},
  \bibinfo{author}{\bibfnamefont{E.}~\bibnamefont{Salumbides}},
  \bibinfo{author}{\bibfnamefont{G.}~\bibnamefont{Dickenson}},
  \bibinfo{author}{\bibfnamefont{K.}~\bibnamefont{Eikema}}, \bibnamefont{and}
  \bibinfo{author}{\bibfnamefont{W.}~\bibnamefont{Ubachs}},
  \bibinfo{journal}{Journal of Molecular Spectroscopy}
  \textbf{\bibinfo{volume}{300}}, \bibinfo{pages}{44} (\bibinfo{year}{2014}),
  ISSN \bibinfo{issn}{0022-2852}, \bibinfo{note}{spectroscopic Tests of
  Fundamental Physics},
  \urlprefix\url{https://www.sciencedirect.com/science/article/pii/S0022285214000630}.

\bibitem[{\citenamefont{Dickenson et~al.}(2013)\citenamefont{Dickenson, Niu,
  Salumbides, Komasa, Eikema, Pachucki, and Ubachs}}]{H2_10_b}
\bibinfo{author}{\bibfnamefont{G.~D.} \bibnamefont{Dickenson}},
  \bibinfo{author}{\bibfnamefont{M.~L.} \bibnamefont{Niu}},
  \bibinfo{author}{\bibfnamefont{E.~J.} \bibnamefont{Salumbides}},
  \bibinfo{author}{\bibfnamefont{J.}~\bibnamefont{Komasa}},
  \bibinfo{author}{\bibfnamefont{K.~S.~E.} \bibnamefont{Eikema}},
  \bibinfo{author}{\bibfnamefont{K.}~\bibnamefont{Pachucki}}, \bibnamefont{and}
  \bibinfo{author}{\bibfnamefont{W.}~\bibnamefont{Ubachs}},
  \bibinfo{journal}{Phys. Rev. Lett.} \textbf{\bibinfo{volume}{110}},
  \bibinfo{pages}{193601} (\bibinfo{year}{2013}),
  \urlprefix\url{https://link.aps.org/doi/10.1103/PhysRevLett.110.193601}.

\bibitem[{\citenamefont{Liu et~al.}(2009)\citenamefont{Liu, Salumbides,
  Hollenstein, Koelemeij, Eikema, Ubachs, and Merkt}}]{H2_D0}
\bibinfo{author}{\bibfnamefont{J.}~\bibnamefont{Liu}},
  \bibinfo{author}{\bibfnamefont{E.~J.} \bibnamefont{Salumbides}},
  \bibinfo{author}{\bibfnamefont{U.}~\bibnamefont{Hollenstein}},
  \bibinfo{author}{\bibfnamefont{J.~C.~J.} \bibnamefont{Koelemeij}},
  \bibinfo{author}{\bibfnamefont{K.~S.~E.} \bibnamefont{Eikema}},
  \bibinfo{author}{\bibfnamefont{W.}~\bibnamefont{Ubachs}}, \bibnamefont{and}
  \bibinfo{author}{\bibfnamefont{F.}~\bibnamefont{Merkt}},
  \bibinfo{journal}{The Journal of Chemical Physics}
  \textbf{\bibinfo{volume}{130}}, \bibinfo{pages}{174306}
  (\bibinfo{year}{2009}), \urlprefix\url{https://doi.org/10.1063/1.3120443}.

\bibitem[{\citenamefont{Liu et~al.}(2010)\citenamefont{Liu, Sprecher, Jungen,
  Ubachs, and Merkt}}]{D2_D0}
\bibinfo{author}{\bibfnamefont{J.}~\bibnamefont{Liu}},
  \bibinfo{author}{\bibfnamefont{D.}~\bibnamefont{Sprecher}},
  \bibinfo{author}{\bibfnamefont{C.}~\bibnamefont{Jungen}},
  \bibinfo{author}{\bibfnamefont{W.}~\bibnamefont{Ubachs}}, \bibnamefont{and}
  \bibinfo{author}{\bibfnamefont{F.}~\bibnamefont{Merkt}},
  \bibinfo{journal}{The Journal of Chemical Physics}
  \textbf{\bibinfo{volume}{132}}, \bibinfo{pages}{154301}
  (\bibinfo{year}{2010}), \urlprefix\url{https://doi.org/10.1063/1.3374426}.

\bibitem[{\citenamefont{Koelemeij et~al.}(2007)\citenamefont{Koelemeij, Roth,
  Wicht, Ernsting, and Schiller}}]{HD_40}
\bibinfo{author}{\bibfnamefont{J.~C.~J.} \bibnamefont{Koelemeij}},
  \bibinfo{author}{\bibfnamefont{B.}~\bibnamefont{Roth}},
  \bibinfo{author}{\bibfnamefont{A.}~\bibnamefont{Wicht}},
  \bibinfo{author}{\bibfnamefont{I.}~\bibnamefont{Ernsting}}, \bibnamefont{and}
  \bibinfo{author}{\bibfnamefont{S.}~\bibnamefont{Schiller}},
  \bibinfo{journal}{Phys. Rev. Lett.} \textbf{\bibinfo{volume}{98}},
  \bibinfo{pages}{173002} (\bibinfo{year}{2007}),
  \urlprefix\url{https://link.aps.org/doi/10.1103/PhysRevLett.98.173002}.

\bibitem[{\citenamefont{Salumbides et~al.}(2013)\citenamefont{Salumbides, Koelemeij,
  Komasa, Pachucki, and Ubachs}}]{Yukawa_Molecule}
\bibinfo{author}{\bibfnamefont{E.~J.}~\bibnamefont{Salumbides}},
  \bibinfo{author}{\bibfnamefont{J.~C.~J.}~\bibnamefont{Koelemeij}},
  \bibinfo{author}{\bibfnamefont{J.}~\bibnamefont{Komasa}},
  \bibinfo{author}{\bibfnamefont{K.~S.~E.}~\bibnamefont{Pachucki}}, \bibnamefont{and}
  \bibinfo{author}{\bibfnamefont{W.}~\bibnamefont{Ubachs}},
  \bibinfo{journal}{Phys. Rev. D.} \textbf{\bibinfo{volume}{87}},
  \bibinfo{pages}{112008} (\bibinfo{year}{2013}),
  \urlprefix\url{https://link.aps.org/doi/10.1103/PhysRevD.87.112008}.
\end{thebibliography}
\end{document}